\documentclass[superscriptaddress,twocolumn,showpacs]{revtex4}
\usepackage{graphicx}
\usepackage{dcolumn}
\usepackage{bm}

\begin{document}

\title{Elongation and fluctuations of semi-flexible polymers in a nematic solvent}

\author{Z. Dogic} \affiliation{Department of Physics and Astronomy,
University of Pennsylvania, Philadelphia, Pennsylvania 19104}

\author{J. Zhang} \affiliation{Department of Physics and Astronomy,
University of Pennsylvania, Philadelphia, Pennsylvania 19104}

\author{A.W.C. Lau} \affiliation{Department of Physics and Astronomy,
University of Pennsylvania, Philadelphia, Pennsylvania 19104}

\author{H. Aranda-Espinoza} \affiliation{Institute for Medicine
and Engineering, University of Pennsylvania, Philadelphia,
Pennsylvania 19104}

\author{P. Dalhaimer} \affiliation{Institute for Medicine
and Engineering, University of Pennsylvania, Philadelphia,
Pennsylvania 19104}

\author{D.E. Discher} \affiliation{Institute for Medicine
and Engineering, University of Pennsylvania, Philadelphia,
Pennsylvania 19104}

\author{P.A. Janmey} \affiliation{Institute for Medicine
and Engineering, University of Pennsylvania, Philadelphia,
Pennsylvania 19104}

\author{Randall D. Kamien} \affiliation{Department of Physics and Astronomy,
University of Pennsylvania, Philadelphia, Pennsylvania 19104}

\author{T.C. Lubensky} \affiliation{Department of Physics and Astronomy,
University of Pennsylvania, Philadelphia, Pennsylvania 19104}

\author{A.G. Yodh} \affiliation{Department of Physics and Astronomy,
University of Pennsylvania, Philadelphia, Pennsylvania 19104}

\date{\today}

\begin{abstract}
We directly visualize single polymers with persistence lengths ranging from
$\ell_p=0.05$ to $16$ $\mu$m, dissolved in the nematic phase of rod-like {\it fd}
virus.  Polymers with sufficiently large persistence length undergo a
coil-rod transition at the isotropic-nematic transition
of the background solvent. We quantitatively analyze the transverse
fluctuations of semi-flexible polymers and show that at long wavelengths
they are driven by the fluctuating nematic background.
We extract both the Odijk deflection length and the elastic constant
of the background nematic phase from the data.
\end{abstract}


\maketitle

\begin{figure}[bp]
{\par\centering
\resizebox{3in}{!}{\rotatebox{0}{\includegraphics{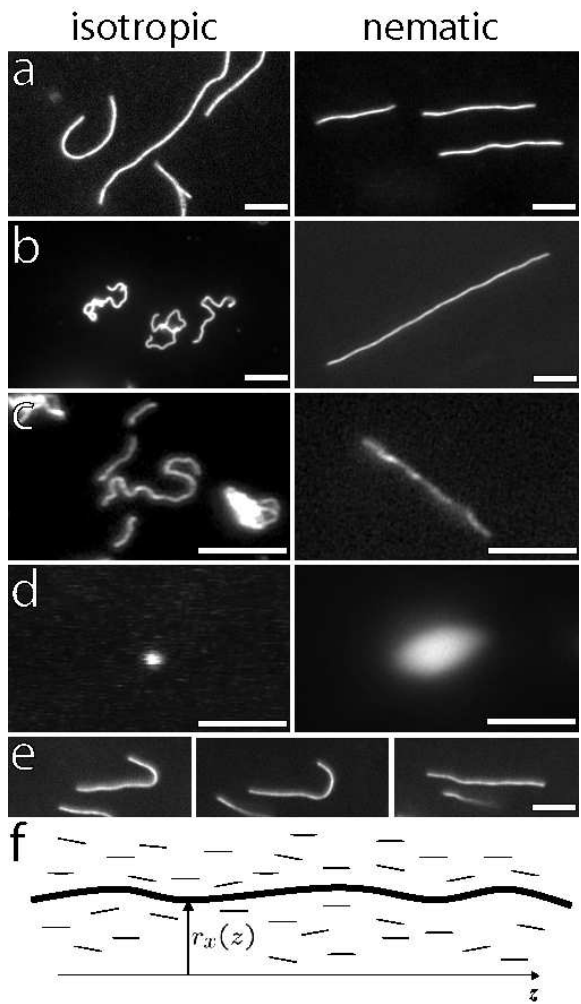}}}
\par}
\caption{Images of fluorescently labelled
biopolymers in the isotropic (left) and nematic (right) phase of {\it fd} virus.
Figures (a)-(d) are, respectively, the images of actin, wormlike micelles, neurofilaments,
and DNA. The polymers in an isotropic solution are confined by a thin chamber thus
making the samples quasi two dimensional.
(e) A sequence of images illustrating an actin filament
escaping from a hairpin defect. The scale bar is 5
$\mu$m. (f) A schematic of a biopolymer in the background
nematic field; the conformation of the polymer is
parameterized by ${\bf R}(z) = \left \{\,r_x(z), r_y(z), z \right \}$.
The nematic director points along the $z$-axis. } \label{figure1}
\end{figure}

Polymer coils in solution exhibit a variety of conformational and
dynamical behaviors depending on many factors, including polymer
concentration, polymer stiffness, solvent quality, solvent flow,
and mechanical stress.  Exciting recent experiments in this field
have focused on disentanglement of single biopolymers in {\em
isotropic} solutions as a result of applied forces and solvent
flow \cite{chu} and on transport of single biopolymers through
networks of barriers \cite{craighead}, which is critically
affected by conformational dynamics of the polymers. In this
Letter, we explore conformations of polymer coils in {\em
anisotropic} solutions. In particular, we present the first direct
experimental observations of isolated semi-flexible polymers
dissolved in a background nematic phase composed of aligned
rod-like macromolecules.  We show by direct visualization that
semi-flexible biopolymers dissolved in the nematic phase assume an
elongated rod-like configuration aligned with the background
nematic director.  The coil-rod transition of the biopolymer can
thus be induced by causing the solvent to undergo an
isotropic-to-nematic (I-N) transition by increasing the
concentration of its constituent rods.  We quantitatively explore
the fluctuations of these semi-flexible polymers and find they
cannot be described by a theory which treats the nematic
background as a fixed external field~\cite{warner}.

Mixtures of semi-flexible polymers in lyotropic nematic
suspensions exemplify an emerging class of complex fluids --
hyper-complex fluids, for example nematic elastomers~\cite{Warner} and
nematic emulsions~\cite{Weitz}, wherein two or more
distinct components are combined to create systems that exhibit novel physical properties
and functions.  Understanding the polymer-nematic system may lead to new ideas about
how to achieve high alignment of biopolymers that is complementary to
existing methods of DNA alignment~\cite{BenNama02}. Furthermore, since many biopolymers
such as the actin filaments within the sarcomere and neurofilaments
within the axon reside in an anisotropic, nematic-like environment~\cite{Aldoroty}, our
investigation may shed light on organization mechanisms within the cell.

We have used fluorescence microscopy to study four different biopolymers in
isotropic and nematic colloidal suspensions.  This approach yields
new information about dynamics and defects not readily accessible to
traditional probes such as x-ray or neutron scattering~\cite{xray}.
In addition, we have developed a rotationally-invariant free energy for a
single semiflexible polymer in a nematic matrix which generalizes
the work in~\cite{deGennes82,Kamien92}, and
enables us to extract the Odijk length~\cite{Odijk86} and the elastic
constant of the liquid crystal.  These first direct measurements
of the Odijk deflection length, $\lambda$, allow us to
quantify the length scale over which the polymer wanders before
it is deflected back by the nematic director.

\begin{table}
\caption{\label{table1} The contour length $L$, the persistence
length $\ell_p$, and the diameter $a$ of bio-polymers in our
experiments.}
\vspace{0.01in}
\begin{tabular}{ccccc}
\hline \hline
 Polymer & $L$ [$\mu$m] & $\ell_p$ [$\mu$m]  & $a$ [nm] & Ref. \\ \hline
 $\lambda$-DNA & 16 & 0.05 & 2 & \cite{Wang97} \\
 neurofilament & 2-10 & 0.2 & 10 & \cite{Aranda03}\\
wormlike micelles & 5-50 & 0.5 & 14 & \cite{Won99} \\
F-actin & 2-20 & 16 & 7 & \cite{actin} \\
{\it fd} virus & 0.9 & 2.2 & 7& \cite{Dogic01} \\
\hline \hline
\end{tabular}
\end{table}

Our experiments employ an aqueous solution of rod-like {\it fd}
viruses as a background nematic liquid crystal.
This system has been studied extensively~\cite{Tang,Dogic01,Purdy03},
and its phase behavior is well described by
the Onsager theory for rods with hard core repulsion~\cite{Onsager49}.
Another advantage of this system is its compatibility with most biopolymers.
We use four different semi-flexible polymers, whose physical parameters
are listed in Table \ref{table1}.  To directly visualize the polymers dissolved
in the nematic background, we fluorescently labelled each polymer:
DNA was labelled with YOYO-1 (Molecular Probes, Eugene OR),
neurofilaments with succinimidyl rhodamine B~\cite{Leterrier96},
F-actin filaments with rhodamine-phalloidin
(Sigma, St.\ Louis MO), and wormlike micelles with PKH26
dye (Sigma, St.\ Louis MO) which preferentially partitions into the
hydrophobic core of the micelle. Since DNA, neurofilaments, and actin are all
negatively charged, we expect that they are stable
in a suspension of negatively charged {\it fd} viruses. Wormlike
micelles are sterically stabilized with a neutral PEO brush layer, which does
not interact with {\it fd} virus or other proteins~\cite{Won99,Dogic01}.

Bacteriophage {\it fd} was grown and dialyzed against  a phosphate
buffer (150 mM KCl, 20 mM phosphate, 2 mM
MgCl$_2$, pH=7.0)~\cite{Dogic01}. Samples were prepared by
mixing a small amount of polymer with {\it fd} solution at different
concentrations and were placed between a coverslip and a glass
slide. A chamber with a thickness of $\sim 50\,\mu$m was made by
using a stretched parafilm as a spacer. Samples sealed with optical glue
(Norland Products, Cranbury, NJ)
were allowed to equilibrate until no drift was visually detectable.
To reduce photobleaching, we added anti-oxygen solution
($2$ mg/ml glucose, $360$ U/ml catalase,
0.25 vol$\%$ mercaptoethanol, 8 U/ml glucose oxidase). All samples were
imaged with a fluorescence microscope (Leica IRBE) equipped with a
100x oil-immersion objective and a $100$ W mercury lamp. Images
were taken with a cooled CCD camera (CoolSnap HQ, Roper
Scientific), which was focused at
least 5 $\mu$m away from the surface to minimize possible wall effects.

\begin{figure}[bp]
{\par\centering
\resizebox*{2.9in}{!}{\rotatebox{0}{\includegraphics{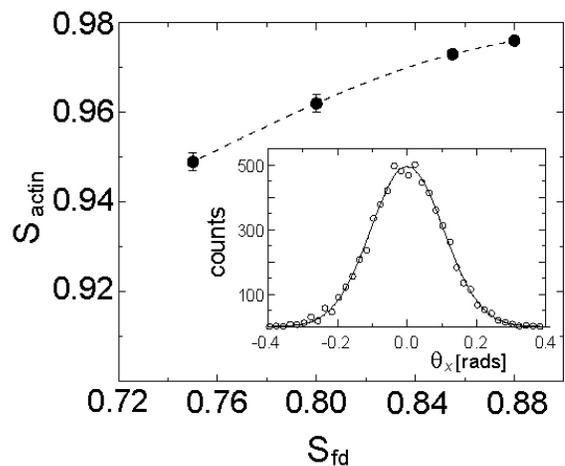}}}
\par}
\caption{The order parameter of actin filaments ($S_{actin}$) vs.\
the order parameter of the background {\it fd} nematic ($S_{fd}$).
Dashed line is a guide to the eye. The contour length of actin
filaments is 15 $\mu$m or higher. The values of $S_{fd}$ at
different {\it fd} concentration have been measured in Ref.\ \cite{Purdy03}.
Inset: The orientational distribution function (ODF) of actin filaments.
The ODF is well approximated by a Gaussian distribution
for a wide range of concentrations.}
\label{figure2}
\end{figure}

Figure~\ref{figure1} displays a series of pictures that summarize
our qualitative observations.  In the nematic phase of $fd$,
F-actin (Fig.~\ref{figure1}a), wormlike micelles (Fig.~\ref{figure1}b),
and neurofilaments (Fig.~\ref{figure1}c) are highly elongated,
having a rod-like shape. By contrast, the same filaments dissolved
in an isotropic phase crumple into more compact random coils. Just
above the I-N transition, actin filaments and worm-like micelles form
hairpin defects~\cite{deGennes82}.
These hairpins exhibit interesting dynamics as shown in Fig.~\ref{figure1}e,
and will be explored by us in detail elsewhere. DNA dissolved
in {\it fd} nematic behaves qualitatively differently (Fig.~\ref{figure1}d); it forms a
slightly anisotropic droplet.
Each droplet contains many DNA molecules and, with time,
these droplets coalesce into a larger droplet.
Thus, even at a very low concentration, DNA
separates from the {\it fd} nematic.  Taken together,
these observations suggest that the
persistence length of the polymer is important in determining its
solubility in the nematic liquid crystals: DNA
has a small $\ell_p$ and is insoluble, unlike the other stiffer polymers
in our experiments.  This insolubility may be related to the
entropy-driven phase-separation of a system of bidisperse
rigid rods if their lengths and/or diameters are
sufficiently dissimilar~\cite{Roij96c}.
This theory, however, has not been extended to the case of semi-flexible polymers.

The large contour lengths of actin
filaments and wormlike micelles make them suitable for further
quantitative analysis.  We focus on the fluctuations of filaments in a background
nematic that is free of both defects and distortions. A series of 50 to
100 images where taken with a few seconds between each image to ensure
that statistically independent configurations were sampled. For
each {\it fd} concentration, ten filaments were analyzed. The conformation
of each polymer was reconstructed by manually marking the end points.
(Note that we parameterize the transverse deviations of the polymer
from the $z$ axis by the 2-component vector ${\bf r}(z)$,
as shown in Fig.~\ref{figure1}f.)  An intensity profile along
the $x$ direction for each value of $z$ was extracted. By fitting this
intensity to a Gaussian, we obtained sub-pixel
accuracy for $r_x(z)$.  We first extracted the orientational distribution function (ODF)
from our data. Since our images are two dimensional projections of the polymer fluctuating in
three dimensions, the $x$-component of the tangent vector is measured by $t_x(z) = \partial\,r_x(z)/ \partial z
\approx \theta_x(z)$.  The ODF is obtained by creating a histogram of $\theta_x$ at
different positions along the contour length for a time sequence of 50-100 images. A
typical ODF is plotted in Fig.~\ref{figure2}a; it is well
approximated by a Gaussian distribution.

\begin{figure}[bp]
\resizebox*{2.9in}{!}{\rotatebox{0}{\includegraphics{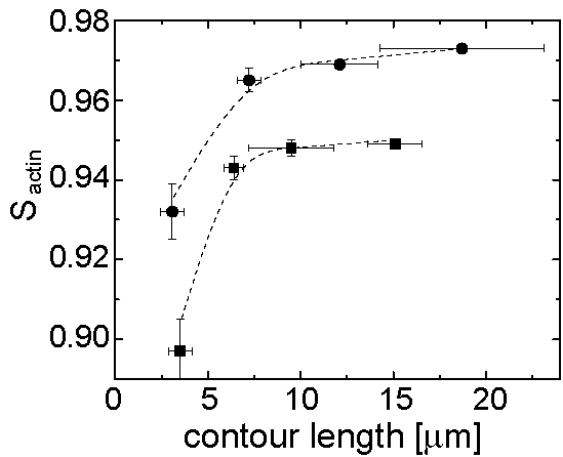}}}
\caption{$S_{actin}$ vs.\ contour lengths of actin.
The concentrations of the background nematic {\it
fd} are 41 mg/ml (squares, $S_{fd}=0.75$) and 28 mg/ml (circles,
$S_{fd}=0.855$). Dashed lines are a guide to the eye.}
\label{figure3}
\end{figure}

Next, we compute the order parameter of the polymer defined by:
$ S \equiv \int_0^{L}\,dz\,\langle \, 3\,( {\bf t}(z)\cdot \hat{z})^2 - 1 \, \rangle /(2L) =
1 - 3 \,\langle \,t_x^2(0) \,\rangle,$
where in the last equality, we have used $\langle \,t_x^2(z) \,\rangle = \langle \,t_y^2(z) \,\rangle$.
In Fig.~\ref{figure2}, we plot $S$ for actin as a function of
the background nematic order parameter.  It is interesting to observe
that $S_{actin}$ is significantly higher than $S_{fd}$.  In order to check that
the difference in the alignment between actin and {\it fd} molecules
is due to their different contour lengths, we measured $S_{actin}$ for different contour
lengths of actin filaments, as shown in Fig.~\ref{figure3}.
As the contour length of actin decreases,
$S_{actin}$ approaches $S_{fd}$, as expected intuitively.
These observations are qualitatively consistent with
the Onsager theory for a bidisperse mixture of rod-like particles with different
lengths considered in Ref.~\cite{Lekkerkerker84}.  This theory predicts the order
parameter of long rods will be higher than the order parameter of the background nematic of
shorter rods.

\begin{figure}[bp]
\resizebox*{3in}{!}{\rotatebox{0}{\includegraphics{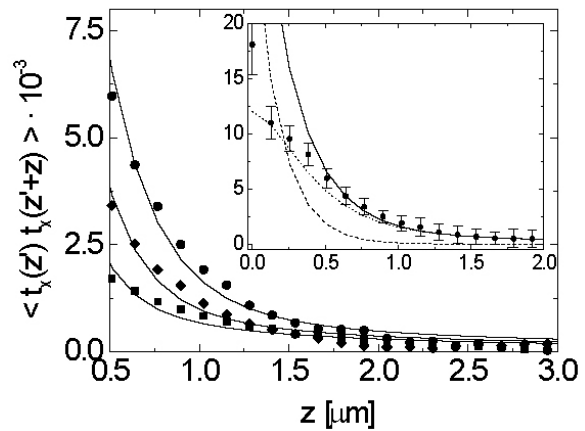}}}
\caption{The $x$ component of the tangent-tangent
correlation function for wormlike micelles measured at three
different {\it fd} concentrations ($c_{fd}$). With increasing {\it fd}
concentration the magnitude of the correlation function decreases.
The solid lines are theoretical curves generated from Eq.\ (\ref{eq:four})
with the best-fit parameters listed in Table~\ref{table2}.
Inset: TTCF for the lowest concentration of the {\it fd} virus.
The dashed and dotted lines are respectively the contributions of the
first and second term in Eq.\ (\ref{eq:four}).  The data points below 0.5 $\mu$m
are unreliable and have been excluded from the fitting~\cite{fudge}.}
\label{figure4}
\end{figure}

Finally, we measured the
tangent-tangent correlation function (TTCF)
$\langle\,t_x(z')\,t_x(z'+z)\,\rangle$
for wormlike micelles dissolved in  {\it fd}
virus with concentration 40 mg/ml and above (Fig.~\ref{figure4}). At low {\it
fd} concentrations, the fluctuations of worms are large as
evidenced by visual observation of spontaneous formation and
dissolution of hairpin defects. In this regime, the
measured TTCF does not decay uniformly. We thus focus our analysis on the regime where
the background order parameter is very high, and the
amplitude of the polymer fluctuations is small. This makes our data
suitable for comparison with the theoretical model outlined below.

The fluctuations of a semi-flexible polymer in a nematic phase may be described
by the free energy~\cite{Kamien92,Selinger}:
\begin{eqnarray}
F & = & {k_B T\over 2} \int^L_0 \,dz \left \{ \ell_p  \left
[\,\frac{\partial {\bf t}_{\bot}} {\partial z} \, \right ]^2  +
\Gamma \left [\, {\bf t}_{\bot}(z) - \delta {\bf n} (0, z)\,
\right ]^2 \right \}\nonumber \\
 &+ & \frac{1}{2} \int d^3x\,K\,({\nabla} \delta {\bf n} )^2,
\label{eq:two}
\end{eqnarray}
where $k_B$ is the Boltzmann constant, $T$ is the
temperature, $\ell_p$ is the persistence length of the semi-flexible
polymer, $\Gamma$ is the strength of the coupling of the polymer
to the background nematic field, $\delta {\bf n}$ is the local
direction of the fluctuating nematic field, and $K$ is the nematic
elastic constant. Note that ${\bf t}_{\bot}$ and $\delta {\bf n}$
are two-dimensional vectors in the plane perpendicular to the average director.
It is straightforward to compute
$\langle\,t_x(z')\, t_x(z'+ z)\,\rangle$ from Eq.\ (\ref{eq:two}):
\begin{eqnarray} \label{eq:four} &&
\langle\,t_x(z')\, t_x(z'+ z)\,\rangle = \frac{\lambda}{4
\ell_p}\,e^{-z/\lambda}
+  \frac{1}{8 \pi^2 K \lambda}  \\
&& \times \int_{0}^{\infty}dx \,
\frac{ \cos (x z/\lambda) \log(1+{D^2/x^2})} {(1+x^2)\left [ 1 + x^2 +
\frac{\Gamma x^2}{4 \pi K}\,\log(1 + {D^2/x^2}) \right ]},   \nonumber
\end{eqnarray}
where $D = 2 \pi \lambda /a $ is related to the molecular cutoff which we assume to
be the diameter of the polymer ($a \sim 10$ nm).  The first term in Eq.\ (\ref{eq:four}) describes
the fluctuations of a semi-flexible polymer in a static external field, with
a decaying length set by $\lambda=\sqrt {\ell_p/\Gamma}$.  The second term describes
the fluctuations of the polymers driven by the tight coupling to the fluctuations of the
background nematic field.  Note that this term generalizes that of Ref.~\cite{Kamien92},
in that it includes the back reaction of the stiff polymer on the nematic fluctuations.
Since it decays approximately as a power law, we expect at large lengthscales the
fluctuations of the polymer are always dominated by the nematic fluctuations.

Figure \ref{figure4} shows our measured TTCF
along with the fitted curve of Eq.\ (\ref{eq:four}).
Overall, good agreement is obtained at distances
above 0.5 $\mu$m~\cite{fudge}.  At these distances,
most of the fluctuations of the worms are driven by the tight coupling
to the background nematic field, coming from the second term
in Eq.~(\ref{eq:four}).  The best-fit value of $\ell_p$
is found to be $1.5\, \mu$m, somewhat higher than that obtained in previous
measurements~\cite{Won99}.  From the fits to the data, we extract
the values of the Odijk deflection length $\lambda$, $K$, and $\Gamma$,
as listed in Table~\ref{table2}.  We observe that with increasing {\it fd}
concentration, $\lambda$ decreases, while $K$ and $\Gamma$ increase,
as one would intuitively expect.  Finally, we note that the values for $K$ are
in agreement with previous measurements of twist elastic constant
$K_{22}=3\cdot10^{-8}$ dyne for {\it fd} samples prepared under
similar conditions~\cite{Dogic00c}.

\begin{table}
\caption{\label{table2} The Odijk deflection length $\lambda$, the elastic
constant of the background nematic $K$, and the coupling constant
between wormlike micelles and background nematic $\Gamma$ for
different {\it fd} concentrations obtained from the fits shown in Fig.~\ref{figure4}.
The best-fit value of $\ell_p$ of wormlike micelles is 1.5 $\mu m$.}
\vspace{0.01in}
\begin{tabular}{cccc}
\hline \hline
 $c_{fd}$ [mg/ml] & $\lambda$ [$\mu$m]  & $K$  [$10^{-8}$ dyne] & $\Gamma$ [1/$\mu$m]\\
  \hline
  39 & 0.18 & 1.9 & 46  \\
 51 & 0.13 & 2.4 & 88  \\
97 & 0.06 & 2.8 & 416  \\
 \hline \hline
\end{tabular}
\end{table}

In conclusion, we have shown that semi-flexible polymers with large
enough persistence lengths assume a rod-like conformation when dissolved
in a nematic solvent. Using image analysis, a full nematic
orientational distribution function was measured. In addition, we have
shown that fluctuations of the polymer are driven primarily
by the fluctuations of the background nematic field.
Direct visualization of individual polymer yields valuable new information
about the behavior of polymer chains in anisotropic solvents.

This work was supported by the NSF through grant DMR-0203378 (AGY),
DMR01-29804 (RDK), and the MRSEC Grant DMR-0079909, and from
NASA NAG8-2172 (AGY), the NIH R01 HL67286 (PAJ), and the Donors of the
Petroleum Research Fund, administered by the American Chemical Society (RDK).


\begin{thebibliography}{0}
\bibitem{fudge}
At distances smaller than 0.5 $\mu$m, we observe a significant deviation of our data
from the theoretical curve, which likely arises from the
limited spatiotemporal resolution of our microscope.
Over the image acquisition time of 250 ms, the fast fluctuations
of the polymers at short wavelengths are effectively washed out, leading
a lower value in the TTCF.  Another possible source of discrepancy is
that the spatial resolution of the microscope is smaller than the
Odijk deflection length.
\end{thebibliography}
\end{document}